\documentclass[sigconf]{acmart}

\usepackage{color}

\usepackage[utf8]{inputenc}
\usepackage{subfig} 

\usepackage{amssymb}

\usepackage{graphicx}
\graphicspath{{./figures/}}

\usepackage{listings}

\usepackage{ifpdf}

\usepackage{textcomp}
\renewcommand{\textrightarrow}{$\rightarrow$}

\usepackage{algorithm, algorithmic}

\ifpdf
\DeclareGraphicsExtensions{.pdf, .jpg, .tif}
\else
\DeclareGraphicsExtensions{.eps, .jpg}
\fi

\setcopyright{acmcopyright}
\begin{document}

\title{A Privacy-Preserving, Accountable and Spam-Resilient Geo-Marketplace}

\author{Kien Nguyen}
\email{kien.nguyen@usc.edu}
\affiliation{%
	\institution{University of Southern California}}
\author{Gabriel Ghinita}
\email{gabriel.ghinita@umb.edu}
\affiliation{%
	\institution{University of Massachusetts Boston}}
\author{Muhammad Naveed}
\email{mnaveed@usc.edu}
\affiliation{%
	\institution{University of Southern California}}
\author{Cyrus Shahabi}
\email{shahabi@usc.edu}
\affiliation{%
	\institution{University of Southern California}}

\copyrightyear{2019} 
\acmYear{2019} 
\acmConference[SIGSPATIAL '19]{27th ACM SIGSPATIAL International Conference on Advances in Geographic Information Systems}{November 5--8, 2019}{Chicago, IL, USA}
\acmBooktitle{27th ACM SIGSPATIAL International Conference on Advances in Geographic Information Systems (SIGSPATIAL '19), November 5--8, 2019, Chicago, IL, USA}
\acmPrice{15.00}
\acmDOI{10.1145/3347146.3359072}
\acmISBN{978-1-4503-6909-1/19/11}



\thispagestyle{empty}
\begin{abstract}

Mobile devices with rich features 
can record videos, traffic parameters or air quality readings along user trajectories. 
Although such data may be valuable, users are seldom rewarded for collecting them.
Emerging digital marketplaces allow {\em owners} to advertise their data to interested {\em buyers}. We focus on {\em geo-marketplaces}, where buyers search data based on geo-tags. 
Such marketplaces present significant challenges. First, if owners upload data with revealed geo-tags, they expose themselves to serious privacy risks. Second, owners must be accountable for advertised data, and must not be allowed to subsequently alter geo-tags. 
Third, such a system may be vulnerable to intensive spam activities, where dishonest owners flood the system with fake advertisements.
We propose a geo-marketplace that addresses all these concerns. We employ searchable encryption, digital commitments, and blockchain to protect the location privacy of owners while at the same time incorporating accountability and spam-resilience mechanisms. We implement a prototype with two alternative designs that obtain distinct trade-offs between trust assumptions and performance. Our experiments on real location data show that one can achieve the above design goals with practical performance and reasonable financial overhead.

\end{abstract}

\maketitle

\renewcommand{\shortauthors}{}
\renewcommand{\shorttitle}{}

\section{Introduction}
\label{sec:intro}

The mobile computing landscape is witnessing an unprecedented number of devices that can acquire geo-tagged data, e.g., mobile phones, wearable sensors, in-vehicle dashcams, and IoT sensors. 
These devices, owned by a diverse set of entities, can collect large amounts of data such as images, videos, movement parameters, or  environmental measurements. 
The data may be useful to third-party entities interested in gathering information from a certain location.
For example, journalists may want to gather images around an event of interest for their newspaper;
law enforcement may seek images taken soon before or after a crime occurred; 
and city authorities may be interested in travel patterns during heavy traffic.

Currently, data collected by individuals are often discarded or archived, due to lack of storage space. Even when data are shared, owners are seldom rewarded for their contributions.
An emerging trend is to create data marketplaces where {\em owners} advertise their data {\em objects} to potential {\em buyers}. 
We emphasize that marketplaces differ from crowdsourcing services such as Amazon Mechanical Turk.
In crowdsourcing, data are owned by the service provider, and the user receives a small reward for a task, e.g., a few cents for classifying an image. In contrast, with data marketplaces users {\em own} the data and advertise them to buyers. If an object is appealing (e.g., a photo purchased by a newspaper), the buyer may pay a higher price (in the order of tens of dollars or more), resulting in different cost and scalability considerations.

Geo-marketplaces, where entities trade geo-tagged data objects, 
raise unique concerns. 
Publishing geo-tags in clear reveals owners' whereabouts, which may lead to serious privacy breaches such as leakage of one's health status or political orientation.
In addition, one must also protect the interests of {\em buyers}, and ensure they receive data objects satisfying their spatial requirements.
Owners must be held {\em accountable} for their advertised data and not be able to change the geo-tag of an object after its initial advertisement. 
This can prevent situations where owners change geo-tags to reflect ongoing trends in buyers' interest. For example, when a certain high-profile event occurs at a location, dishonest owners may attempt to change their geo-tags closer to that location in order to sell their images at higher prices. Furthermore, the system must provide strong disincentives to prevent {\em spam} behavior, where dishonest participants flood the system with fake advertisements.

We propose a geo-marketplace with three key features: 

\noindent
{\bf Privacy.} We adapt state-of-the-art {\em searchable encryption (SE)} techniques to protect locations, and we perform matching between buyer interests and advertised objects on {\em encrypted} geo-tags. 

\noindent
{\bf Accountability.} To hold owners accountable for their advertisements, we use {\em cryptographic commitments} and {\em blockchain} technology. We store a compact digital commitment on the blockchain to prevent owners from altering object geo-tags after publication. 

\noindent
{\bf Spam-Resilience.} We employ the use of a public blockchain, where writing to the ledger requires a transaction fee. We control the cost such that legitimate users only pay negligible fees relative to the value of their objects, whereas dishonest users who flood the system with fake advertisements are strongly disincentivized.

In our design, the data owner generates a metadata item which includes the object's geo-tag. The {\em bulk} data (e.g., image or video), is either stored by the owner (e.g., flash drive), or encrypted with conventional encryption at a bulk storage service, such as Swarm~\cite{swarm} or InterPlanet File System~\cite{ipfs}. The low-footprint geo-tag metadata is encrypted using SE. The owner then creates a {\em digital commitment} of the metadata and stores it on the blockchain. 
Commitments can be stored either individually, or batched together for better blockchain efficiency and cost. 

Buyers search objects based on geo-tags by querying the encrypted metadata. They must first obtain a {\em search token} that allows them to identify encrypted objects that match their spatial range query. Since processing on encrypted data is computationally expensive, if the buyers decide to use other services to perform the task, they often need to pay for the search token and its processing. In our system model, different strategies are investigated to ensure that the performance and financial cost of the search process are practical. It is important to remark that the encrypted search reveals neither the exact whereabouts of the objects, nor the owner's identity. The buyer learns only pseudonymous owner identifiers for matching objects, e.g., a blockchain public key, through which the transaction can be anonymously completed.
Once matching objects are identified, the owner and buyer enter a {\em smart contract} through the blockchain. As a result, the owner receives payment, and the buyer receives the actual data objects, and the corresponding conventional decryption keys.

Achieving the three aforementioned objectives is challenging. 
First, SE techniques incur significant overhead compared to the search on plaintexts, especially with asymmetric encryption. Thus, carefully designing data and query encodings is essential to obtain efficient solutions that can be scaled to large datasets. 
Second, the cost of privacy and accountability should not be too high; otherwise, it may interfere with the financial operation of the marketplace, resulting in prohibitive costs. An acceptable financial cost should only account for a small percentage of the transaction value. 

Our specific contributions include:

\begin{itemize}
	\vspace{-5pt}
	\item
	We propose a novel architecture for a geo-marketplace that achieves privacy, accountability, and spam resilience by combining searchable encryption, digital commitments, and block-chain. To the best of our knowledge, this is the first work aiming to accomplish these objectives.
	
	\item
	We propose protocols for owner-buyer matching with both symmetric and asymmetric SE. These approaches offer an interesting trade-off between trust assumptions and performance, facilitating adoption in a wide range of scenarios.
	
	\item
	We develop optimization techniques to address the high computational cost of encrypted search. We also consider techniques to decrease the financial cost of blockchain operations by reducing the amount of on-chain storage.
	
	\item
	We perform an extensive experimental evaluation to measure system performance, in terms of computational overhead, storage, and financial cost incurred.
	\vspace{-5pt}
\end{itemize}

Sec.~\ref{sec:background} provides background information on the different components of the system, followed by an overview of the system model and operations workflow in Sec.~\ref{sec:systemmodel}. 
We present technical details in Sec.~\ref{sec:approach} and experimental results in Sec.~\ref{sec:experiments}. We review related work in Sec.~\ref{sec:relatedwork} and conclude with directions for future research in Sec.~\ref{sec:conclusion}.

\vspace{-5pt}
\section{Background}
\label{sec:background}

\noindent
{\bf Searchable Symmetric Encryption (SSE)} allows a client to search and selectively
retrieve her encrypted documents outsourced to a server. SSE was first proposed
in~\cite{song2000practical} and further refined in~\cite{goh2003secure,curtmola2011searchable}. 
The first efficient
sub-linear SSE scheme that supports Boolean queries was proposed in~\cite{cash2013highly}.
Later on,~\cite{stefanov2014practical} proposed a scheme that achieves forward security by protecting access patterns at the time of document addition.
State-of-the-art SSE schemes are efficient, 
but at the expense of some leakage in the form of access patterns. 
In our system, we use the recently-proposed HXT technique~\cite{HXT18} which supports conjunctive keyword queries.

Let $d_0, \dots, d_{n-1}$ be the client's
documents and $I$ an inverted index that maps a keyword $w$ to the list
of document identifiers containing $w$. We denote the
list of document identifiers that contain $w$ as $I(w)$. An SSE scheme consists
of the following four algorithms:

\noindent
{\em 1) Setup} is run by the client and takes as input security parameter $k$ and documents $d_0, \dots, d_{n-1}$.
		It generates two secret keys $K_I$ and $K_D$.  It parses all the documents
		and forms an inverted index $I$ that maps to each keyword $w$ a list of
		document identifiers ($I(w)$) that contain $w$. The client encrypts this index
		using a special encryption algorithm specified by the particular SSE scheme
		and generates an encrypted inverted index using the key $K_I$. It also
		encrypts each document $d_i, \forall 0 \leq i < n$, with conventional symmetric
		encryption (e.g., AES) using the key $K_D$ and assigns it a unique identifier
		that is independent of the document contents.  It outputs the keys $K_I$ and
		$K_D$ that are stored locally at the client and the encrypted index 
		that is sent to the server.

\noindent
{\em 2) Token Generation}, run by the client, takes as input secret key $K_I$ and a keyword $w$. Using secret key $K_I$, it creates a search token $tk_w$ which is sent to the server.

\noindent
{\em 3) Search} is run by the server and uses as input the token $tk_w$
		and the encrypted index. It searches the encrypted index and retrieves
		the list of document identifiers that contain the keyword $w$, namely
		$I(w)$. The server retrieves the encrypted documents $d^w_0, \dots,
		d^w_{|I(w)|-1}$ using the identifiers in $I(w)$ and sends the documents to
		the client.

		\noindent
{\em 4) Decryption} is run by the client and uses the secret key $K_D$ to decrypt the
		documents received from the server.

\noindent
{\bf Hidden Vector Encryption (HVE)}~\cite{Boneh06,Boneh07} is an {\em asymmetric} searchable encryption technique supporting conjunctive equality, range and subset queries. 
Search on ciphertexts can be performed with respect to a number of {\em index attributes}. HVE represents an attribute as a bit vector (each element has value $0$ or $1$), and the search predicate as a {\em pattern} vector where each element can be $0$, $1$ or '*' (i.e., wildcard value). Let $l$ denote the HVE {\em width}, which is the bit length of the attribute, and consequently that of the search predicate. A predicate evaluates to $True$ for a ciphertext $C$ if the attribute vector $I$ used to encrypt $C$ has the same values as the pattern vector of the predicate in all positions that are not '*' in the latter.

HVE is built on top of a symmetric bilinear map of composite order \cite{Boneh05}, which is a function $e : \mathbb{G} \times \mathbb{G} \rightarrow \mathbb{G}_T$ such that $\forall a,b \in G$ and $ \forall u,v \in \mathbb{Z}$ it holds that $e(a^u,b^v)=e(a,b)^{uv}$. $\mathbb{G}$ and $\mathbb{G}_T$ are cyclic multiplicative groups of composite order $n=p\cdot q$ where $p$ and $q$ are large primes of equal bit length. We emphasize that the application of function $e$, which is called a {\em bilinear pairing}, is expensive to compute, so the number of pairings must be minimized.
We denote by $\mathbb{G}_p$, $\mathbb{G}_q$ the subgroups of $\mathbb{G}$ of orders $p$ and $q$, respectively. HVE consists of the following four algorithms:

\noindent
{\em 1) Setup.} The private/public key pair ($SK$/$PK)$ are as follows:
$$SK = ( g_q \in \mathbb{G}_q,\quad a \in \mathbb{Z}_p,\quad \forall i \in [1..l]: u_i,h_i, w_i, g, v \in \mathbb{G}_p )$$
$$PK = (g_q,\quad V=vR_v,\quad A=e(g,v)^a,\quad$$
$$\forall i \in [1..l]: U_i=u_iR_{u,i},\quad  H_i=h_iR_{h,i},\quad  W_i=w_iR_{w,i})$$
with random \(R_{u,i}, R_{h,i}\), \(R_{w,i} \in \mathbb{G}_q, \forall i \in [1..l]\)  and \(R_v \in \mathbb{G}_q\)

\noindent
{\em 2) Encryption} uses $PK$ and takes as parameters index attribute $I$ and message $M \in \mathbb{G}_T$. The following random elements are generated: \(Z, Z_{i,1}, Z_{i,2} \in \mathbb{G}_q\) and \(s \in \mathbb{Z}_n\). The ciphertext is: 
$$C = (C^{'}= MA^s,\quad C_0=V^sZ, \quad $$
$$\forall i \in [1..l]: C_{i,1} = (U^{I_i}_iH_i)^sZ_{i,1}, \quad C_{i,2} = W^{s}_iZ_{i,2} )$$

\noindent
{\em Token Generation.} Using $SK$, and given a search predicate encoded as pattern vector \(I_{*}\), the TA generates 
a search token $TK$ as follows: let \(J\) be the set of all indices $i$ where \(I_{*}[i] \neq *\).
TA randomly generates \(r_{i,1}\) and \(r_{i,2} \in \mathbb{Z}_p, \forall i \in J\). 
Then
$$TK=(I_*, K_0 = g^a\prod_{i \in J}(u^{I_{*}[i]}_ih_i)^{r_{i,1}}w^{r_{i,2}}_i, \quad$$
$$ \forall i \in [1..l]: K_{i,1} = v^{r_i,1},\quad K_{i,2} = v^{r_i,2})$$

\noindent{\em Query} is executed at the server, and evaluates if the predicate represented by $TK$ holds for ciphertext $C$. The server attempts to determine the value of \(M\) as 
\begin{equation}
M = C^{'}{/} (e(C_0,K_0) {/} \prod_{i \in J} e(C_{i,1},K_{i,1}) e(C_{i,2},K_{i,2}) \label{eq:query}
\end{equation}
If the index $I$ on which $C$ was computed satisfies $TK$, the value of \(M\) is returned, otherwise a nil value $\bot$ is obtained.

\noindent
{\bf Vector Digital Commitments.}
Cryptographic commitments~\cite{PedersenCommitment91} allow a party $S$ to commit to a message $m$ by
creating a commitment $CC$,  such  that  $CC$ is binding (i.e., $S$ cannot change
the  message $m$) and hiding (i.e., $CC$ does not leak any information about $m$).
In this work, we use vector
commitments~\cite{catalano2013vector}, which allow party $S$ to commit to an
ordered sequence of messages $(m_0, \dots, m_{q-1})$, such that it can  later
open the commitment for a specific message, e.g., to prove that $m_i$  is the
$i$-th message in  the sequence. Vector commitments are space-efficient because 
their size is
independent of the number of committed values. 
 A vector commitment scheme is defined by the following four
algorithms:

\noindent
{\em 1) KeyGen} takes as input security
		parameter $k$ and size $q$ of committed vector and outputs a public
		parameter $pp$.

\noindent
{\em 2) Commit}  takes  as input a sequence of $q$ messages $V = m_0, \dots,
		m_{q-1}$, the public parameter $pp$ and outputs a commitment string $CC$ and
		an auxiliary information $aux$.

\noindent
{\em 3) Open} takes as input a message $m \in {m_0, \dots, m_{q-1}}$, a
		position $i$, and the  auxiliary information $aux$ and is run by  the
		committer to produce a proof $P_i$ that $m$ is the $i$-th message
		in the committed message vector $V$.

\noindent
{\em 4) Verify} takes as input the commitment string $CC$, a message $m$,
		a  position $i$, and the proof $P_i$, to verify that $P_i$ is a valid proof
		that $CC$  was created to a  sequence $m_0, \dots, m_{q-1}$, where $m = m_i$.

\noindent
{\bf Blockchain, Smart Contracts and Bulk Storage.} Blockchain was first introduced in~\cite{nakamoto2008bitcoin} as a decentralized public ledger that records transactions among entities without a trusted party. %
A blockchain is a sequence of transaction blocks cryptographically-linked through the hash value of the predecessor.
A transaction typically moves cryptocurrency from one account to another.
An account is defined as the public key of an entity, which provides pseudonymity. 
System nodes called {\em miners} compete to create new blocks by solving {\em proof-of-work} puzzles. 
The miner who finds the puzzle solution first is rewarded with cryptocurrency. 

Some blockchain platforms (e.g., Ethereum), have the ability to execute smart contracts~\cite{szabo1996smart}, which are sophisticated agreements among entities that utilize transactions on the blockchain. Smart contracts are expressed in a high-level programming language (e.g., Solidity) interpreted by a blockchain virtual machine. 
One limitation when storing data on the blockchain is size. Due to the competitive nature of block creation, the growth rate of the blockchain is limited. 
Recently, decentralized storage systems have been proposed that interface with the blockchain and allow large amounts of storage (e.g., Swarm~\cite{swarm}). Such systems provide a {\em distributed hash table (DHT)} interface~\cite{Chord} to store and retrieve data. Participating peers receive incentives for the contributed storage.

\vspace{-8pt}
\section{System Model}\label{sec:systemmodel}

The central component in our design is the blockchain, and its associated {\em on-chain} operations. On-chain storage is financially expensive, since {\em write} operations to the chain translate into transaction blocks added to the ledger. We aim to minimize the amount of on-chain storage. Only digital commitments and minimal addressing information is stored on-chain. For all other data structures, we employ bulk storage (Swarm).
Another challenging part of the system is matching owners' data to buyers' requests. This process involves search on encrypted location metadata, which is computationally expensive, especially in the case of asymmetric searchable encryption. Searchable ciphertexts tend to be large in size compared to corresponding plaintexts, in order to support conjunctive queries and hide data patterns. In the case of SSE, metadata ciphertexts and associated indexes are also placed in bulk storage.

We present evaluation metrics in Section~\ref{sec:metrics}, followed by two alternative system designs: in Section~\ref{sec:sse-arch} we present a solution based on SSE, which achieves sub-linear search performance, thanks to the use of an encrypted index. However, this approach requires a {\em trusted curator (TC)}, which holds the secret encryption key, and has access to the plaintext locations of all object geo-tags. In Section~\ref{sec:hve-arch}, we propose an asymmetric encryption design, where each owner has the public key of a private/public key pair. Owners encrypt locations using Hidden Vector Encryption (HVE). There is still need for a {\em trusted authority} (TA) that holds the private key and generates search tokens at runtime, but this entity does {\em not} have access to plaintext locations. While in principle it is possible for the TA to collude with buyers and issue numerous search tokens that may reveal all object locations, such an attack is more difficult to stage. We assume that the TA is non-colluding. In addition, it is possible to use multiple TAs, so the amount of disclosure in the case of collusion is limited. The disadvantage of HVE is that it does not allow the construction of an index, so a linear search is required. %

\vspace{-8pt}
\subsection{Evaluation Metrics}
\label{sec:metrics}

We consider computation time, storage size and financial cost as performance metrics. The latter is measured in Ethereum using the concept of {\em gas}. Each on-chain transaction requires spending a certain amount of gas to complete. The cost of one unit of gas is linked to the Ethereum price, which is market-driven. One unit of gas costs 1e-9 $ether$. At the time of writing, $1ether=\$133$ . Since on-chain operations are dominated by the cost of gas, which far exceeds the corresponding storage and computational overhead, we exclusively use gas to measure on-chain operations.
For off-chain operations, we use computation time to evaluate the performance of: location encryption and index generation (indexes are used only for SSE), cryptographic commitment generation, search token generation and token-object matching. In terms of storage cost, we focus on ciphertext size and search token size. Commitment size is taken into account using gas, as it is stored on-chain.

\begin{figure}[tb]
	\begin{center}
		\includegraphics[width=1.0\columnwidth]{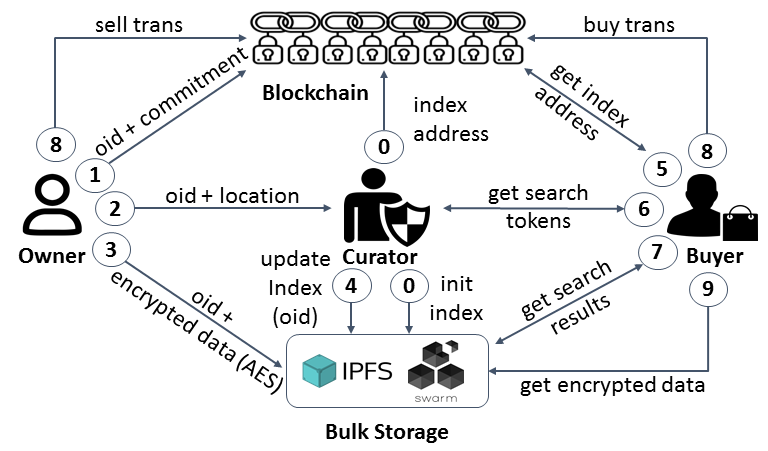}
		\vspace{-20pt}
		\caption{SSE-based System Workflow}
		\label{fig:sse-model}  
		\vspace{-20pt}
	\end{center}
\end{figure}

\vspace{-5pt}
\subsection{Private Geo-marketplace with SSE}
\label{sec:sse-arch}

We employ HXT encryption~\cite{HXT18}, the state-of-the-art in conjunctive keyword search. HXT builds an index which allows sub-linear search. 
Fig.~\ref{fig:sse-model} shows the system architecture, with three types of entities: {\em owners}, {\em buyers} and a {\em trusted curator} (TC). TC collects plaintext locations of all objects and builds an index with SSE-encrypted locations. In practice, the curator role can be fulfilled by an entity that already has an established relationship of trust with owners, e.g., a cell operator that already has access to customer locations. The TC has financial incentives to operate the service: it can charge a small fee for location indexing and search token generation. The system can have more than one curator, each serving a subset of owners. While this reduces the amount of location exposure to a single entity, it results in multiple indexes, which may reduce search efficiency. In the rest of the paper, we assume a single curator.

The TC initializes the index (Step 0 in Fig.~\ref{fig:sse-model}), assigns it a unique index identifier (IID) and stores it in bulk storage. The IID is also stored on the blockchain, and later used by buyers to bootstrap the search. For example, the IID may include TC contact information such as URI.
Each data owner is represented by a pseudonymous identifier, e.g., public key in the blockchain system. To advertise an object, the owner randomly generates a unique object identifier (OID) and computes a digital commitment that covers the geo-tag and OID (Step 1). 
The digital commitment is stored on the blockchain.
Since the commitment is hiding, placing it on the blockchain will not disclose the object's location. The binding property of commitments, combined with the unmalleable storage property of the blockchain, guarantees that the owner can be held accountable if it turns out that the advertised object is collected at another location than the advertised one. 
Next (Step 2), the owner submits the plaintext location along with the OID to the TC, and uploads the bulk data (encrypted with conventional AES) to bulk storage (Step 3) having the OID as key (recall that, the bulk storage offers a DHT-like interface and stores key-value pairs). 
Then, the curator inserts the object in the encrypted HXT index (Step 4).
We emphasize that location proofs are orthogonal to our approach, and existing solutions~\cite{locationproof-BrambillaAZ16,foam,platin,xyo} can be adopted in our system.

Buyers search objects based on geo-tags. We assume buyers specify search predicates in the form of spatial range queries.
The buyer locates the TC bootstrap information (Step 5), and contacts the TC with the search predicate in plaintext. The TC, who holds the master secret key of the SE  instance, generates a search token to evaluate the spatial predicate. The token is sent to the buyer (Step 6). The curator may charge the buyer a fee for each token. Flexible pricing policies may be implemented by the curator: for instance, tokens that cover a larger area, or that cover denser areas where one is expected to find more objects, may be more expensive.

Next (Step 7), the actual search is performed using the HXT index. The index is stored in distributed fashion on bulk storage, so the search process can be completed by the buyer through repeated interaction with the DHT interface using the IID and index pointers as request keys. Alternatively, the buyer can employ another service that performs the search directly on the storage nodes. The details of this process are orthogonal to our approach, and we consider as performance metric the total computational cost incurred by the search, which is the same whether it is executed on the buyer's machine (e.g., in the case of an institutional buyer with significant resources), or on the Swarm nodes (e.g., in the case of a private buyer who performs the transaction using its mobile phone and pays an additional fee for the search).%

When search completes, the buyer learns the pseudo-identities of matching owners, and decides which data object(s) to purchase. The purchase is completed through a smart contract between the owner and the buyer (Step 8), following which the owner receives payment, and the buyer receives the AES key used to encrypt the object in Swarm. The buyer downloads the object (Step 9) and decrypts it locally, at which point the transaction is finalized. If it turns out that the data does not satisfy the advertised geospatial attributes, the buyer can contest the transaction, and use the digital commitment to prove that the owner is dishonest. The payment is reversed, and additional punitive measures (e.g., reputation penalties) can be taken against the owner. The transactions on the blockchain are evidence of the purchase and provide accountability.

\begin{figure}[tb]
	\begin{center}
		\includegraphics[width=1.0\columnwidth]{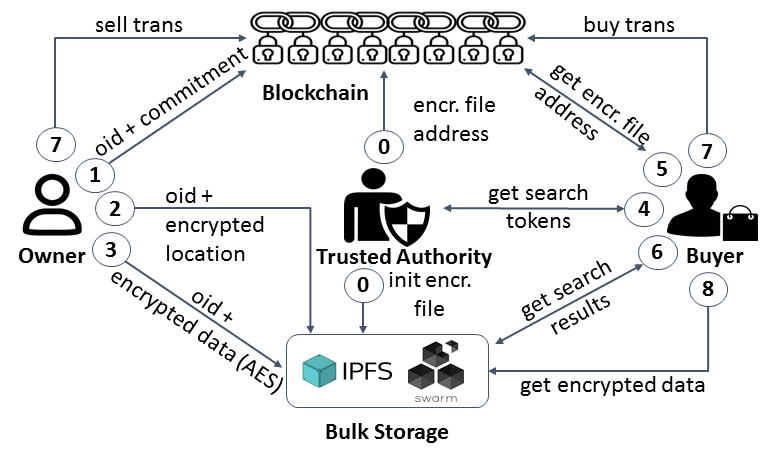}
		\vspace{-20pt}
		\caption{HVE-based System Workflow}
		\label{fig:hve-model}  
		\vspace{-20pt}
	\end{center}
\end{figure}

\subsection{Private Geo-marketplace with HVE}
\label{sec:hve-arch}

Fig.~\ref{fig:hve-model} illustrates the system workflow when using HVE encryption. The main difference compared to the SSE case is the absence of the curator. Instead, there is a Trusted Authority (TA), which holds the private (or master) key of the HVE instance and issues search tokens.
Most of the steps remain the same, with a few exceptions. In Step 0, instead of building an index, the TA initializes a flat file that will contain all the encrypted object locations. In Step 2, the owners encrypt object locations by themselves, using the public HVE key. This reduces considerably disclosure compared to the SSE case. The rest of the workflow remains unchanged. However, the counter of Steps 4-8 in Fig.~\ref{fig:hve-model} is less by one compared to their SSE counterparts, due to the absence of the index update step.

\vspace{-5pt}
\section{Technical Approach}
\label{sec:approach}

Sec.~\ref{sec:ssetechnical} discussed search using SSE, whereas Sec.~\ref{sec:hvetechnical} focuses on HVE. Sec.~\ref{sec:commitments} discusses accountability and SPAM resilience.

\vspace{-5pt}
\subsection{Symmetric Encryption Search}
\label{sec:ssetechnical}
	
SSE techniques~\cite{HXT18,stefanov2014practical} support keyword (i.e., {\em exact match}) queries, and conjunctions thereof, over an arbitrary domain. Our objective is to support {\em range} queries on top of geospatial data. For simplicity, we consider a two-dimensional (2D) space (although our results can be easily generalized to 3D). SSE schemes assume a database of documents, where each document is associated with a set of keywords. In our setting, each object is a document, and its keywords are derived based on the geo-tag of the object. 

\begin{figure}
	\centering
	\includegraphics[width=0.85 \columnwidth]{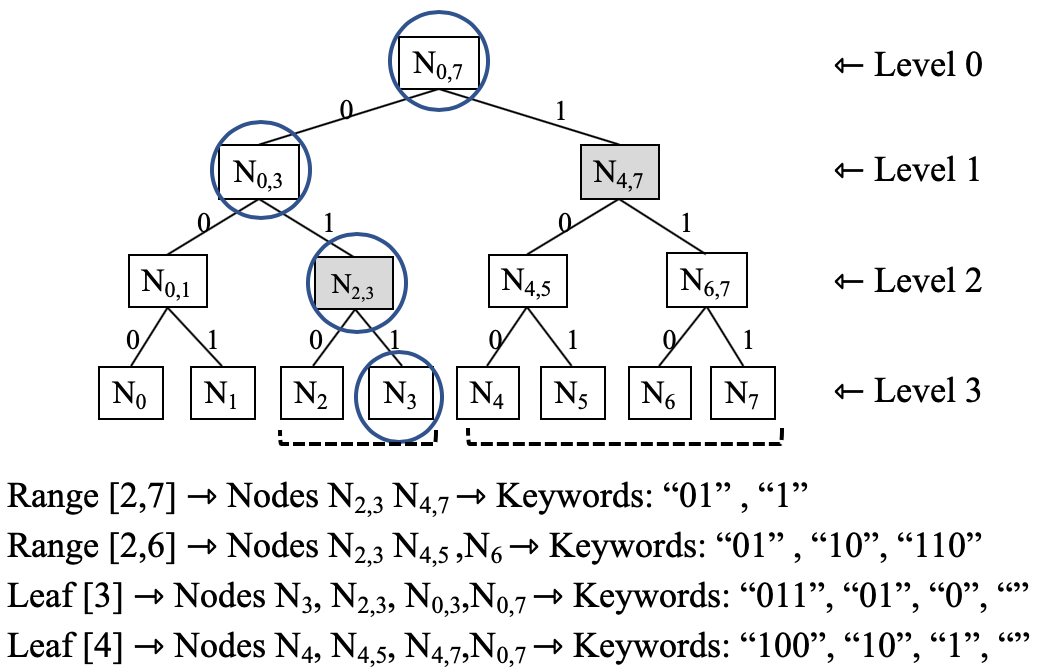}
	\vspace{-10pt}
	\caption{Mapping 1D domain using \textit{best range cover}}

	\label{fig:brc1d}  
	\vspace{-5pt}
\end{figure}

First, we discuss the case of data and range queries for a one-dimensional (1D) domain. Consider a domain $\mathbf{A}$ of integers from $0$ to $L-1$, where $L$ is a power of $2$ (i.e., the domain of $(\log L)$-bit integers). Even though spatial coordinates are real numbers, we can represent them using an integer-valued domain with good precision.
We construct a full binary tree over domain $\mathbf{A}$, where the value in each node represents a domain range.
Each node can be uniquely identified using the path leading to it from the root node. 
Along that path, left branches are labeled with $0$ and right branches with $1$. 
A node identifier ({\em id}) is the unique string that concatenates all edge labels on the path from the root to that node. 
Figure~\ref{fig:brc1d} shows the resulting binary tree for a 3-bit domain $\mathbf{A} = [0, \dots, 7]$. For example, the node id of $N_{2,3}$ in Fig.~\ref{fig:brc1d} is $\text{"01"}$.

We adopt a domain encoding called \textit{best range cover} (BRC)~\cite{Kiayias13}.
Given a range $r$, BRC selects the minimal set of nodes that cover $r$. 
The ids of the nodes in this set represent the \textit{keywords} associated to $r$.
For example, the range $[2, 7]$ is minimally covered by nodes $N_{2,3}$ and $N_{4,7}$ (shown shaded in Fig.~\ref{fig:brc1d}), with node ids (i.e., keywords) $\text{"01"}$ and $ \text{"1"}$; whereas the range $[2, 6]$ is minimally covered by nodes $N_{2,3}, N_{4,5}$ and $N_6$, with node ids $\text{"01"}$, $\text{"10"}$ and $ \text{"110"}$. 
The ranges covering a leaf node can be identified by traversing the tree {\em upward} from that leaf node. Given a data value, its keywords are represented by the node ids from the upward traversal that starts at the leaf node representing the value's binary representation.
For instance, leaf node $N_3$ is covered by nodes $N_{3}$, $N_{2,3}$, $N_{0,3}$, and $N_{0,7}$ (encircled in Fig.~\ref{fig:brc1d}), with node ids $\text{"011"}, \text{"01"}, \text{"0"}$ and $ \text{"$\emptyset$"}$ as keywords (the root node is encoded as $\emptyset$ since its path has length $0$).

For a 2D domain, a separate binary tree is constructed for each dimension, and the node ids are prefixed with $\text{"x"}$ and $\text{"y"}$, respectively, to distinguish values in each coordinate. 
For example, the ids of node $N_{2,3}^x$ and $N_{2,3}^y$ on the tree of \textit{Ox} and \textit{Oy} are represented as $\text{"x01"}$ and $\text{"y01"}$, respectively.
Each location is a single cell in the grid of size $L \times L$ covering the entire geospatial domain.
For an object positioned at cell $(i, j)$ corresponding to leaf nodes $N_i^x, N_j^y$ in their respective domain trees, the union of node ids covering $N_i^x, N_j^y$ is used as the keyword set for the object. The id of the root is omitted since it appears in every object.
The total number of keywords of each object is $2 \log L$.
For instance, the keyword set of an object in cell $(3, 4)$ contains $2 \log 8 = 6$ nodes: $N_{3}^x$, $N_{2,3}^x$, $N_{0,3}^x$, $N_{4}^y$, $N_{4,5}^y$, $N_{4,7}^y$, with labels "x011", "x01", "x0", "y100", "y10", "y1". %

A range query in the 2D domain is a cross join of node ids in each dimension. For example, the 2D range query defined by $x \in [2, 7]$ and $y \in [2, 6]$ is expressed 
as $(N_{2,3}^x \wedge N_{2,3}^y) \vee (N_{2,3}^x \wedge N_{2,3}^y) \vee \dots \vee (N_{2,7}^x \wedge N_{6}^y) $ 
or in our specific encoding $(\text{"x01"} \wedge \text{"y01"}) \vee (\text{"x01"} \wedge \text{"y10"}) \vee \dots \vee (\text{"x1"} \wedge \text{"y110"})$.
Each term in the expression is a \textit{conjunctive keyword query}, which is directly supported by HXT.
HXT query time depends on the number of documents containing the first keyword~\cite{HXT18}. We sort the keywords in a query so that the node closest to the leaf level is in the first position, since that node covers a smaller range, hence query cost is decreased.

The trusted curator (TC) divides the data domain into a $L \times L$ grid with an appropriate domain granularity $L$ (e.g., one can choose $L$ such that one unit corresponds to a distance of one meter). Then, using the range covering technique above, TC builds an HXT-encrypted index for all objects. When a buyer initiates the search process, a range query is performed as a series of {\em conjunctive} keyword queries of length two (due to the 2D domain). If any term evaluates to true for an object, the object matches the query.

Algorithm~\ref{alg:get_1d_covering_nodes} shows the pseudocode to obtain covering nodes ids for each object coordinate. Algorithm~\ref{alg:hxt_convert_doc} shows how to generate the object-keyword database $DDB$, which is subsequently encrypted using HXT's {\em Setup} procedure (Section~\ref{sec:background}). {\em Setup} outputs
master key \textit{mk}, public parameters \textit{pub}, and encrypted database \textit{EDB}. Algorithm~\ref{alg:sse.spatialrangequery} shows the pseudocode for answering encrypted spatial range queries on top of {\em EDB}.

\begin{algorithm} 
	\caption{Get 1D Covering Nodes} 
	\label{alg:get_1d_covering_nodes} 
	\begin{algorithmic} 
		\REQUIRE Domain length $L$; Position \textit{pos} $: 0 \leq \textit{pos}  < L$; 
		\ENSURE Node ids covering position \textit{pos} 
		
		\textit{node\_ids} $\leftarrow$ $\emptyset$
		
		\textit{current\_id} $\leftarrow$ ""
		
		\FOR{$i = \log L$ to $0$}
		\STATE append $\text{i}^{\text{th}}$-bit of \textit{pos} to \textit{current\_id}
		\STATE \textit{node\_ids} = \textit{node\_ids} $\cup$ \textit{current\_id}
		\ENDFOR
		
		\RETURN \textit{node\_ids}
	\end{algorithmic}
\end{algorithm}
\begin{algorithm} %
	\caption{Convert Object Locations} %
	\label{alg:hxt_convert_doc} %
	\begin{algorithmic} %
		\REQUIRE Grid length $L$; Location database $\textit{LDB}$ where $\textit{LDB[i]} = (x_i, y_i)$, $\forall i: 0 \leq x_i, y_i < L$;
		\ENSURE Document database \textit{DDB}
		
		\FORALL{object id $i \in \textit{LDB}$} 
		\STATE \textit{x\_words} $\leftarrow$ Get 1D Covering Nodes of $x_i$
		\STATE \textit{y\_words} $\leftarrow$ Get 1D Covering Nodes of $y_i$
		\FORALL{$w$ $\in$ \textit{x\_words}}
		\STATE $\textit{DDB}[i]$ = $\textit{DDB}[i] \cup concatenate('x',w)$
		\ENDFOR
		\FORALL{$w$ $\in$ \textit{y\_words}}
		\STATE $\textit{DDB}[i]$ = $\textit{DDB}[i] \cup concatenate('y',w)$
		\ENDFOR
		\ENDFOR
		\RETURN $\textit{DDB}$
		
	\end{algorithmic}
\end{algorithm}

\begin{algorithm} 
	\caption{EncryptedSpatialRangeQuery} 
	\label{alg:sse.spatialrangequery} 
	\begin{algorithmic} 
		\REQUIRE Grid size $L$; \textit{EDB}; range $R$; HXT parameters \textit{pub}, \textit{mk}  
		\ENSURE Matched object identifiers
		
		\STATE \textit{x\_words} $\leftarrow$ Get1DCover $[\textit{R.bottom\_right}^x,\textit{R.top\_left}^x]$
		\STATE \textit{y\_words} $\leftarrow$ Get1DCover $[\textit{R.bottom\_right}^y,\textit{R.top\_left}^y]$
		
		\textit{matches} $\leftarrow \emptyset$
		\FORALL{pairs $\textit{kq} = (w_x, w_y) \in \textit{x\_words} \times \textit{y\_words}$}
		\STATE \textit{matches} $\leftarrow$ \textit{matches} $\cup$ HXTSearch(\textit{param}, \textit{mk}, \textit{kq}, \textit{EDB})
		\ENDFOR 
		\RETURN \textit{matches} 

	\end{algorithmic}
\end{algorithm}

\noindent
{\bf Limiting Query Size and Placement.} With the proposed technique, a buyer can issue range queries of arbitrary size, shape, and placement. Arbitrary queries are decomposed into a set of queries that are precisely covered by a domain tree node, and a disjunctive expression is formed, where each term is a conjunctive HXT query. Such flexibility can decrease performance, since there may be numerous sub-queries in the decomposition. In practice, query sizes are likely to be small (e.g., no more than $1km^2$ within a city). In addition, one can slightly restrict query placement, requiring that a range aligns precisely with a tree node.

First, we consider limiting {\em query size}. Specifically, if query size is limited to a maximum of $L / 2 ^ {h_{\max}}$ in each dimension, then one needs to consider only nodes up to a level $h_{\max}$ of the domain tree. The larger $h_{\max}$ is, the fewer node levels are considered.
Nodes at higher levels (i.e., $level < h_{\max}$) can be ignored when constructing the keyword set for each object, resulting in smaller ciphertexts and faster processing.
Index creation time is also significantly boosted.

Second, we restrict queries to areas that are precisely covered by one node of each tree. For example, for the 1D domain in Fig.~\ref{fig:brc1d}, although both ranges $[3, 6]$ and $[4, 7]$ have size $4$, range $[3, 6]$ is covered exactly by three nodes $(N_3, N_{4,5}, N_6)$ while range $[4, 7]$ can be covered with a single node $(N_{4,7})$. 
With this alignment restriction, query decomposition is no longer necessary, and each range query can be encoded as a single conjunctive keyword pair.

\vspace{-5pt}	
\subsection{Asymmetric Encryption Search}
\label{sec:hvetechnical}

The SSE approach described previously is quite efficient, but it requires a trusted curator that has access to all plaintext locations of advertised objects. This can lead to excessive disclosure, and in some cases, it may be unrealistic to assume that system users are willing to place so much trust in a centralized component. The HVE-based approach described next uses asymmetric encryption and allows object owners to encrypt locations by themselves. However, this comes at additional performance overhead.

Previous work that focused on location-based queries on top of HVE-encrypted data considered hierarchical or Gray encodings~\cite{Ghinita14}. In this context, each object location is snapped to a $L \times L$ grid. An attribute vector is constructed for each object, which has width $l = 2 \log L$. Next, queries are expressed with respect to groups of neighbor cells with similar encoded values. This may lead to excessive computation time, given that the number of expensive bilinear pairing operations required to evaluate a single token is proportional to the HVE index vector width (as discussed in Section~\ref{sec:background}).
Furthermore, a range query in~\cite{Ghinita14} often requires more than a single token to evaluate, increasing computation time even more.
Another problem with using multiple tokens for a single query is excessive leakage. To improve performance, the work in~\cite{Ghinita14} uses single-cell token aggregation and may end up with several tokens for each query predicate, corresponding to sub-ranges of the query. Based on the individual evaluation for each sub-query, an adversary may pinpoint the object's location to a smaller area than the actual query of the buyer, which may result in significant privacy leakage.

We propose another approach of encoding range queries using HVE. Essentially, our approach first transforms range queries to keyword queries, using a domain mapping similar to the one used for SSE. Then, HVE is used to assemble a ciphertext that allows HVE evaluation of a conjunctive exact match query, one for each spatial dimension.
Recall from Section~\ref{sec:background} that the elements of an attribute vector can take values in $\Sigma_{*} = \mathbb{Z}_m \cup \{*\}$ for an integer $m < p, q$.
Also, a conjunctive formula of length two formed with respect to node identifiers on \textit{Ox} and \textit{Oy} axes represents a 2D range. 
For example, $[N_{2,3}^x, N_{4,5}^y]$ indicates range $[2,3] \times [4,5]$ in 2D domain.

We utilize the domain mapping technique described in Section~\ref{sec:ssetechnical} to encode both the location of an object and a range query. 
Then, for each coordinate of an object's location, there are $\log L$ nodes in the upward path from the leaf node of the coordinate to the root. To capture all covering ranges on both coordinates, we would need to capture within each ciphertext $\log^2 L$ pairs of $(x,y)$ coordinates. Even for moderate domain representation granularities (e.g., $L=65536, \log L=16$), this would result in a significant storage overhead (e.g., $256$ values for the $\log L=16$ setting). In addition to the storage overhead, there is also increased processing time when performing queries, since all pairs are potential candidates for matching. To prevent performance deterioration, in the case of HVE we choose to adopt a further query limitation compared to SSE: specifically, we consider only query ranges with square shape. These queries can still occur at each level of the domain tree, but the range spans in each dimension are equal. As a result, we only need to store $\log L$ pairs of encrypted coordinates in a HVE ciphertext. In fact, when combined with the maximum query size limit discussed for SSE, the overhead decreases to $\log L - h_{max} + 1$.

Even with this additional constraint, buyers are still able to formulate useful queries. If one considers the families of all regular grids with granularity increasing in powers of two super-imposed on top of the data domain, then our encoding still allows a query to express any possible cell within one of these grids. While clearly more restrictive compared to arbitrary queries, the approach achieves a good trade-off between flexibility and performance. 

\begin{figure}
	\centering
	\includegraphics[width=0.88 \columnwidth]{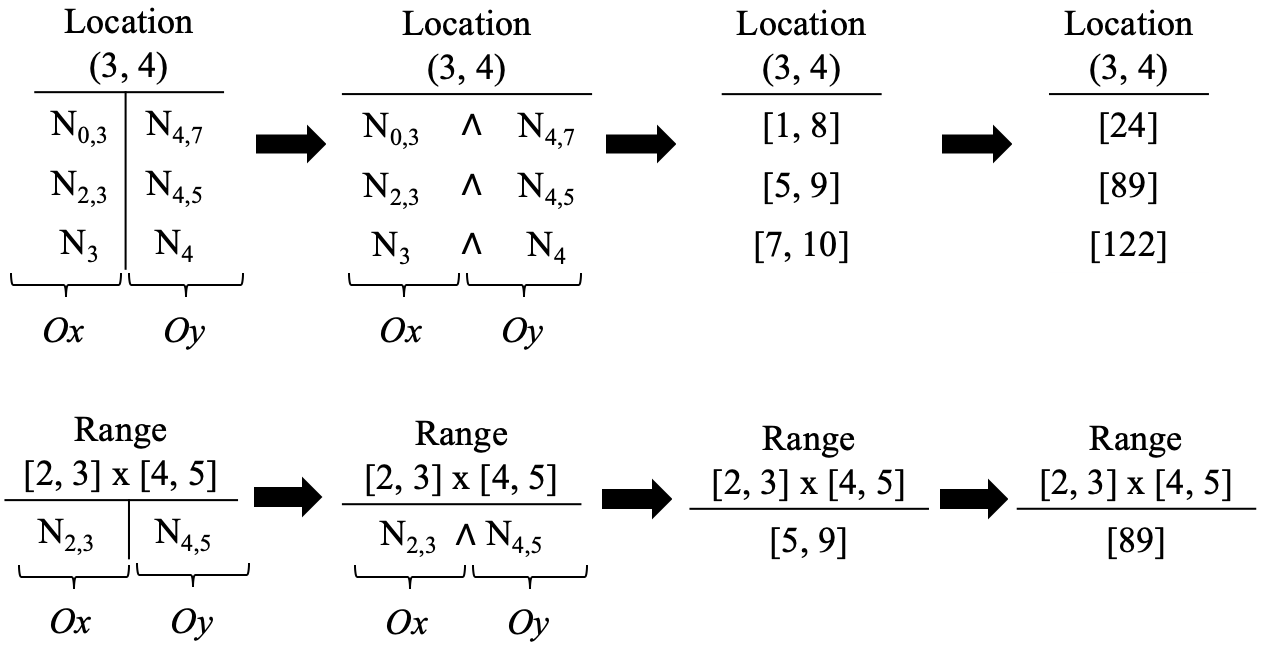}
	\vspace{-10pt}
	\caption{Data and query encoding with HVE}%
	\label{fig:hvepreorder}  
	\vspace{-20pt}
\end{figure}

Fig.~\ref{fig:hvepreorder} exemplifies the modified encoding. For each object, the ids of nodes in the domain binary tree on one dimension are combined with the ids of nodes \textit{at the same level} in the other dimension to form for each ciphertext $\log L$ vectors of length one. Each vector corresponds to a \textit{square} area that is covered exactly by one node of each tree.
Range queries are also restricted in the same fashion, i.e., a \textit{square} area covered exactly by one node of each tree, to allow conjunctive exact match evaluation.
Thus, both the location of an object and a range query are encoded as a single scalar value, which can be used directly in the HVE matching phase. 
In our implementation, the order of the node in a pre-order traversal of the tree, identified by the path from the root to the leaf, is used as the node identifier. 
Then, to ensure each area has a unique value, the final value is calculated as $N^x * 2 * L + N^y$ where $N^x, N^y$ are the node ids of the object in the tree of $Ox$, $Oy$, respectively.
For instance, consider an object at location $(3, 4)$ and a range query $[2, 3] \times [4, 5]$. As shown in Fig.~\ref{fig:hvepreorder}, the location $(3, 4)$ is covered by nodes $N_{0, 3}, N_{2,3}, N_3$ in the tree of \textit{Ox} and $N_{4, 7}, N_{4,5}, N_4$ in the tree of \textit{Oy} (the superscripts are omitted for simplicity). Hence, the ranges covering the object are represented as the combinations of a node in the tree of \textit{Ox} and a node in the same level in the tree of \textit{Oy} (e.g. $[N_{0, 3} \wedge N_{4, 7}], [N_{2, 3} \wedge N_{4,5}], [N_{3} \wedge N_{5}]$). Then, these combinations are transformed into length-two vectors using pre-order traversal of the elements, which are $[1, 8], [5, 9], [7, 10]$. The final calculated values for each case are $[24], [89]$ and $[122]$, respectively. The same encoding is also applied to the range query, resulting in value $[89]$.

Compared to the approach in~\cite{Ghinita14} which results in HVE vectors of length $l = 2\log L$, we generate length-one vectors, where each vector is constructed from nodes of the same level in each tree.
Following the query size limitation, the size of the queries is maximum $L / 2 ^ {h_{\max}}$ in each dimension. 
When a buyer issues a query, the resulting token generated by the TA must be compared with all vectors of the ciphertext, and there is exactly one match if the object is within the query range, and no matches if the object falls outside the range. This results in maximum $\log L - h_{max} + 1$ evaluations. With a simple modification, we can reduce this overhead to a single evaluation, by also including the domain tree level within the query. This way, the matching procedure only needs to evaluate the token against the ciphertext at the same level as the one specified in the query, reducing computational overhead of matching by a factor of $\log L$. Storage requirements for ciphertexts do not change, since there could be a potential query received for any level of the domain tree. We refer to this search variation as {\em SingleLevel}.
	
\vspace{-10pt}
\subsection{Owner Accountability and SPAM Resilience}
\label{sec:commitments}

Two major desiderata of our proposed system, accountability and spam resilience, are achieved by storing for each advertised object a digital commitment on the blockchain. Since write operations to the chain are expensive, and the amount of information that can be stored is small, it is crucial to reduce the size of commitment strings, in order to reduce transaction costs. We employ the use of {\em vector commitments}~\cite{catalano2013vector} which allow the commitment of a sequence of values in a compact way (the length of the commitment string is independent of the number of committed values). This way, an owner can submit at the same time commitments for a batch of objects and pay the on-chain write price for just one. The length of a commitment is equal in size to that of an RSA encryption modulus for a similar amount of security: in our implementation, we use 1024-bit commitments.

To advertise one or more objects, the owner creates a vector digital commitment where each component corresponds to a location. Locations are typically hashed, and then the commitment is created by performing a modulo exponentiation operation with the hashed value in a composite order group. The commitment vector is published on the blockchain. Due to the difficulty of extracting logarithms in the group, an adversary is not able to recover the committed value, so location privacy for the owner is achieved. On the other hand, the owner cannot later change the value of the location without being detected, due to the {\em binding} property of commitments. After a transaction is completed, if the buyer finds that the actual location of the data object differs from the advertised one, the on-chain commitment is sufficient to prove the owner's dishonesty and reverse the payment (additional penalties can be imposed by the marketplace).

Spam-resilience is also achieved as a result of using the blockchain. Due to the non-negligible cost of writing to the blockchain, it is not economically viable for a dishonest party to advertise a large number of objects. Various policies can be put in place to control the trade-off between the cost per transaction incurred by legitimate users, and the resilience to spam. For instance, one can enforce a limit on the count of elements in vector commitments, which in effect determines the number of objects that can be advertised with a fixed cost (in the experimental evaluation of Section~\ref{sec:experiments} we measure the cost of writing a commitment to the chain in the order of USD\$0.02).
In addition, the system can enforce a policy that mandates a deposit for each commitment.
The deposit can be refunded back to the owner after a transaction is completed, or after a pre-defined time threshold.
For example, if the cost is \$0.02 to submit a commitment for $20$ objects (which translates to \$1 for $1000$ objects), the system can enforce an additional deposit of \$1 per commitment (i.e., \$50 for 1000 objects) refundable after one day. This policy can be easily implemented through a smart contract.

Another policy that can be easily implemented as a smart contract is to enforce a hard limit, e.g., $50$ commitments per owner per day. While the deposit requirement policy focuses on the financial aspect, such a thresholding policy directly restricts the number of commitments. Finally, when SSE is used, one can also enforce an anti-spam mechanism at the TC, by limiting the number of objects that are encrypted for each owner. Since the TC will not encrypt more objects than the limit, the index size will be kept under control, and search performance will not be affected.

\section{Experimental Evaluation}
\label{sec:experiments}

\noindent
{\bf Experimental Settings.}
We evaluate the proposed approaches using the SNAP project~\cite{snapgowalla} location dataset, containing check-ins of users in the Gowalla geo-social network. We assume that owner objects' geo-tags coincide with check-in locations.
We select check-ins in the Los Angeles area, spanning latitude range $[33.6996,34.3423]$ North and longitude range $[118.6846,118.1444]$ West (an area of $3500 km^2$ with a total of $110,312$ check-ins). We randomly select from this set four object datasets $D_1, D_2, D_5$ and $D_{10}$ having cardinality $10000, 20000, 50000$ and $100000$, respectively, such that $D_1 \subset D_2 \subset D_5 \subset D_{10}$.
The area is partitioned into a $L \times L$ grid with granularity: $G=\log L = 10$, $12$, $14$ and $16$
(ranging from $2300m^2$ down to less than $1m^2$ per cell).
Object locations were converted to keywords for HXT, and to length-two vectors for HVE.
Search requests of buyers were randomly generated by choosing an anchor location from the dataset, then constructing a range around it with three sizes: $400\mathtt{m} \times 550\mathtt{m}$, $800\mathtt{m} \times 1100\mathtt{m}$, and $1600\mathtt{m} \times 2200\mathtt{m}$, ranging from $1\%$ to $3.5\%$ of data domain side length.  

\begin{figure}
	\subfloat[Index Build Time]{
		\includegraphics[width=0.5 \columnwidth]{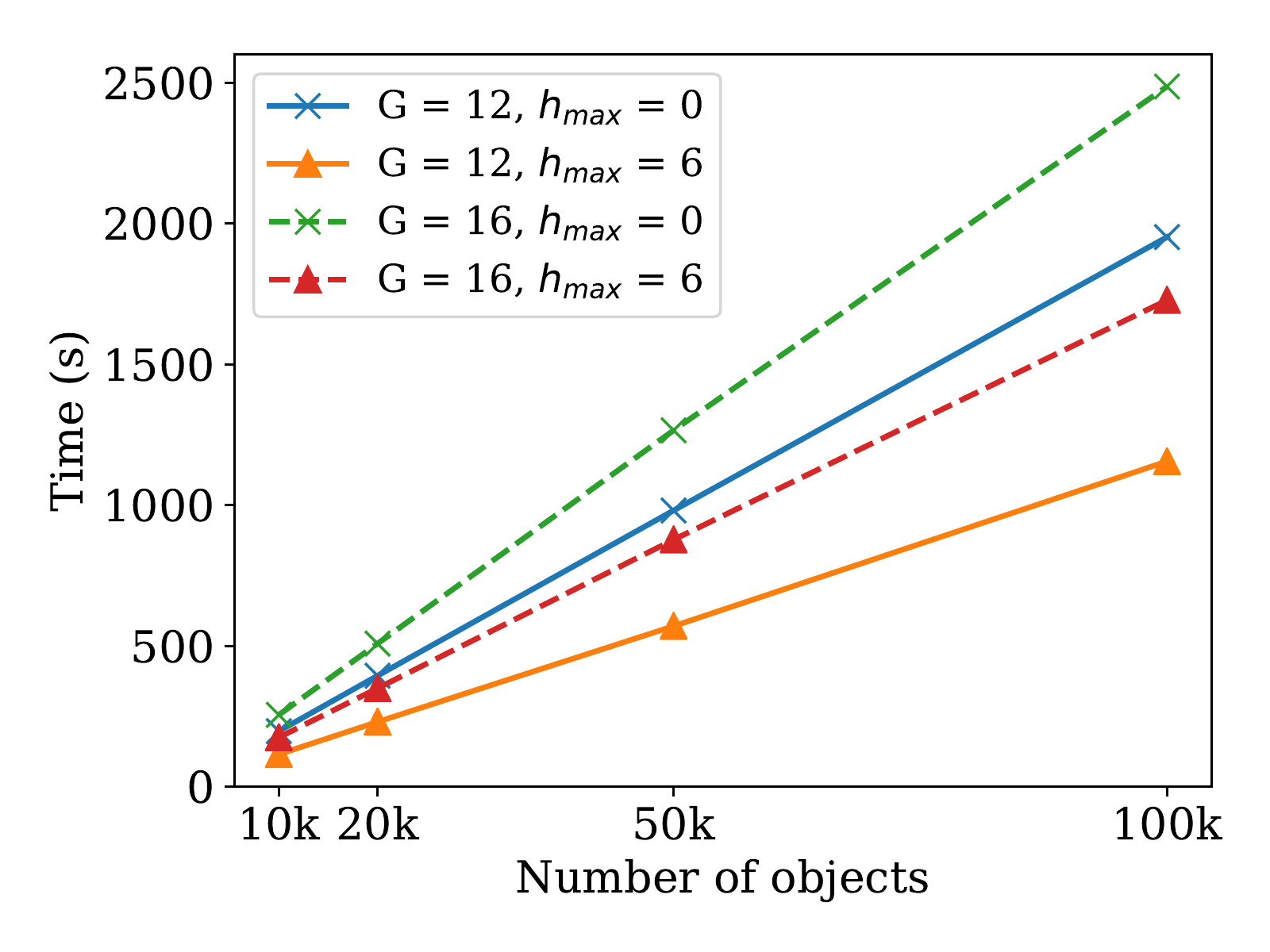}
		\label{fig:hxt_index_build_time}}
	\subfloat[Index Size]{
		\includegraphics[width=0.5 \columnwidth]{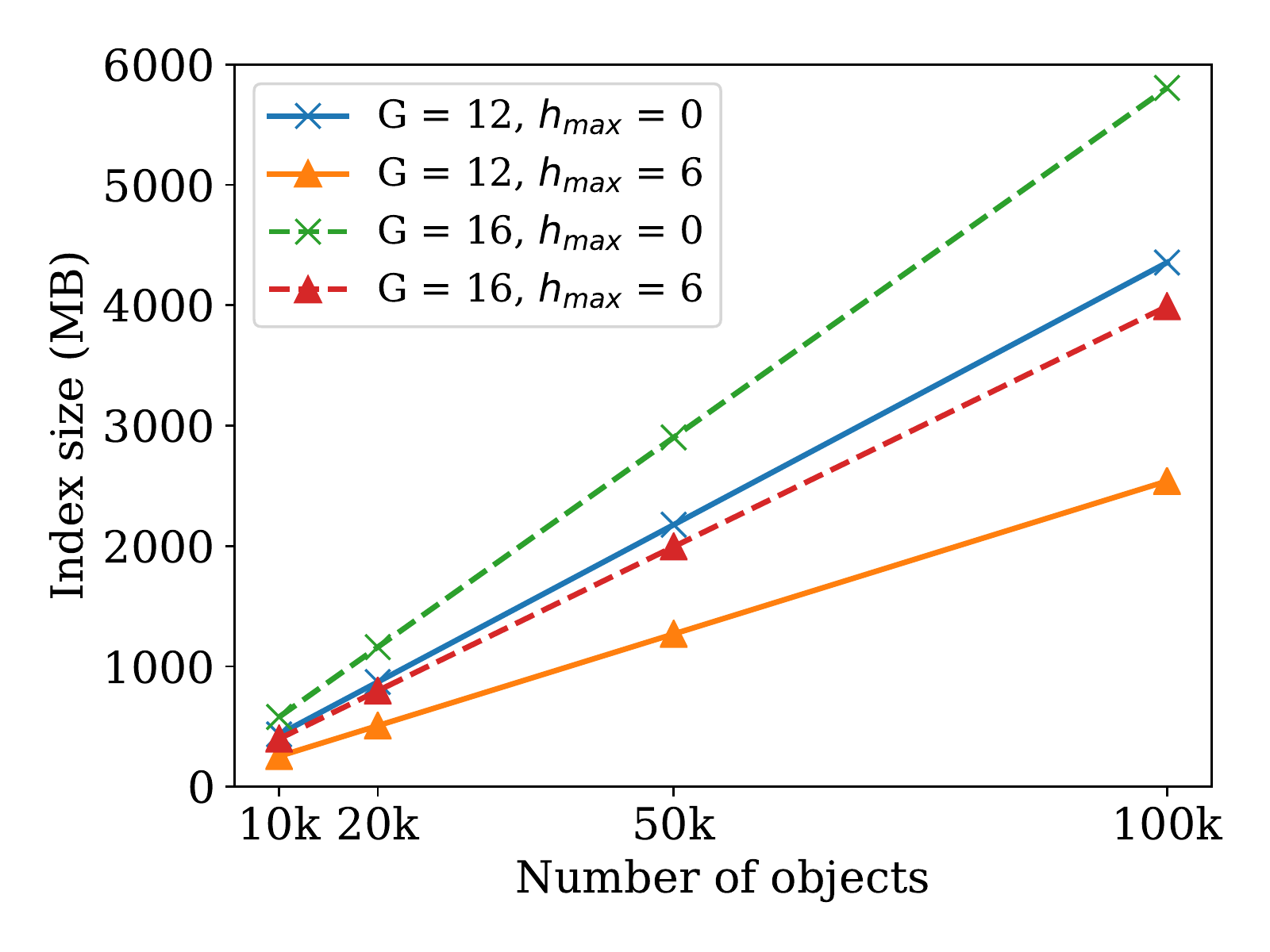}
		\label{fig:hxt_index_size}  }
	\vspace{-5pt}
	\caption{HXT index generation performance}
	\label{fig:hxt_index}
	\vspace{-15pt}
\end{figure}

We implemented Python prototypes of the proposed approaches.
For HXT, we employed 1024-bit key length for pairing groups, while for HVE we used the instantiation from \cite{Boneh07} and varied key length as $768$, $1024$, $1536$ and $2048$ bits.
All experiments were run on a Intel Core i7 3.60GHz CPU with 4 cores and 16GB RAM, running Ubuntu 18.04. We used a single core for all experiments, except the HVE parallel processing test for which we used all four available cores. For blockchain tests, we built a private Ethereum network using Go Ethereum version 1.8.20-stable.%

\begin{figure}
	\subfloat[Arbitrary query placement]{
		\includegraphics[width=0.5 \columnwidth]{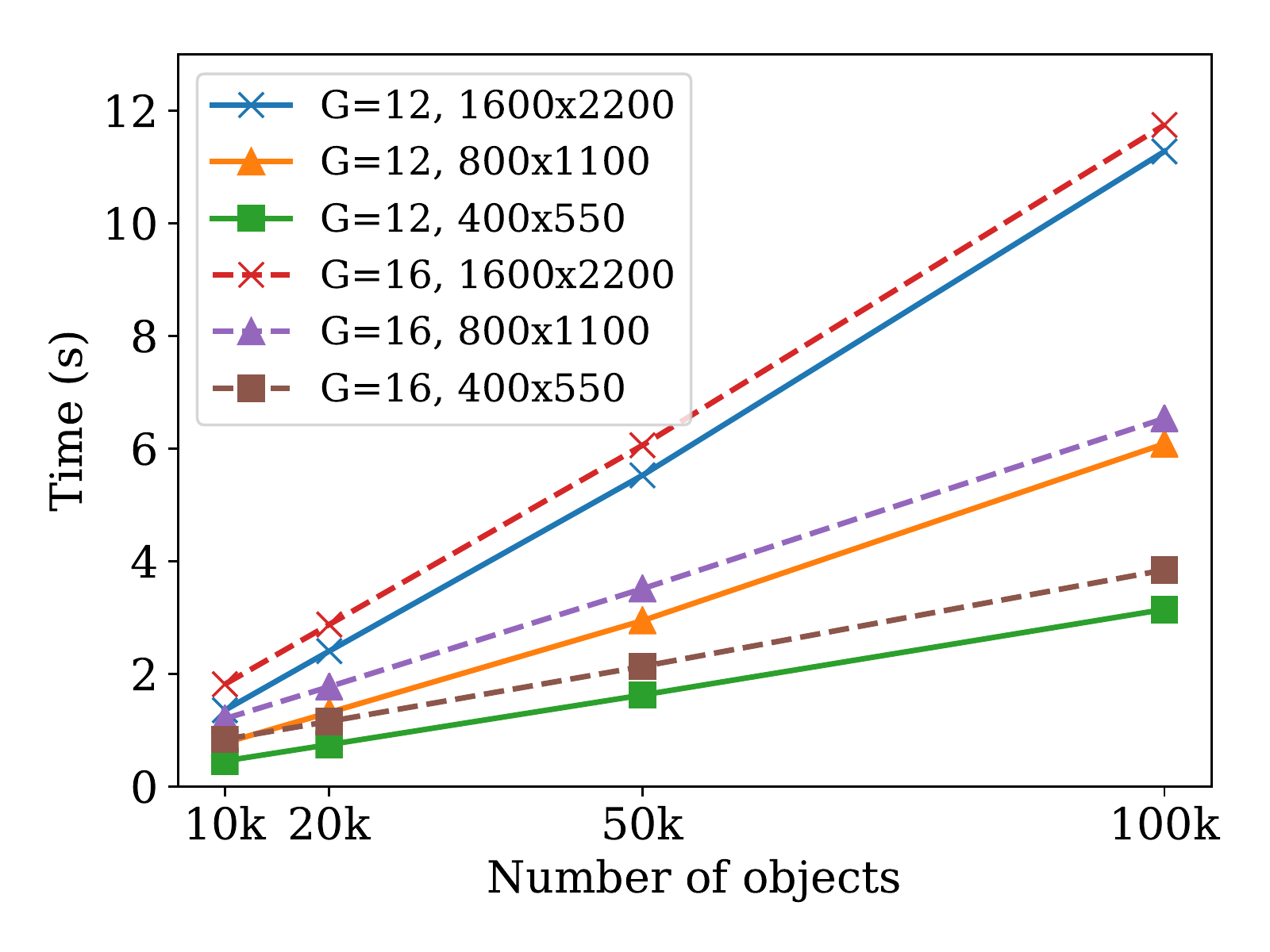}
		\label{fig:hxt_avg_query_time_arbitrary}}
	\subfloat[Restricted query placement]{
		\includegraphics[width=0.5 \columnwidth]{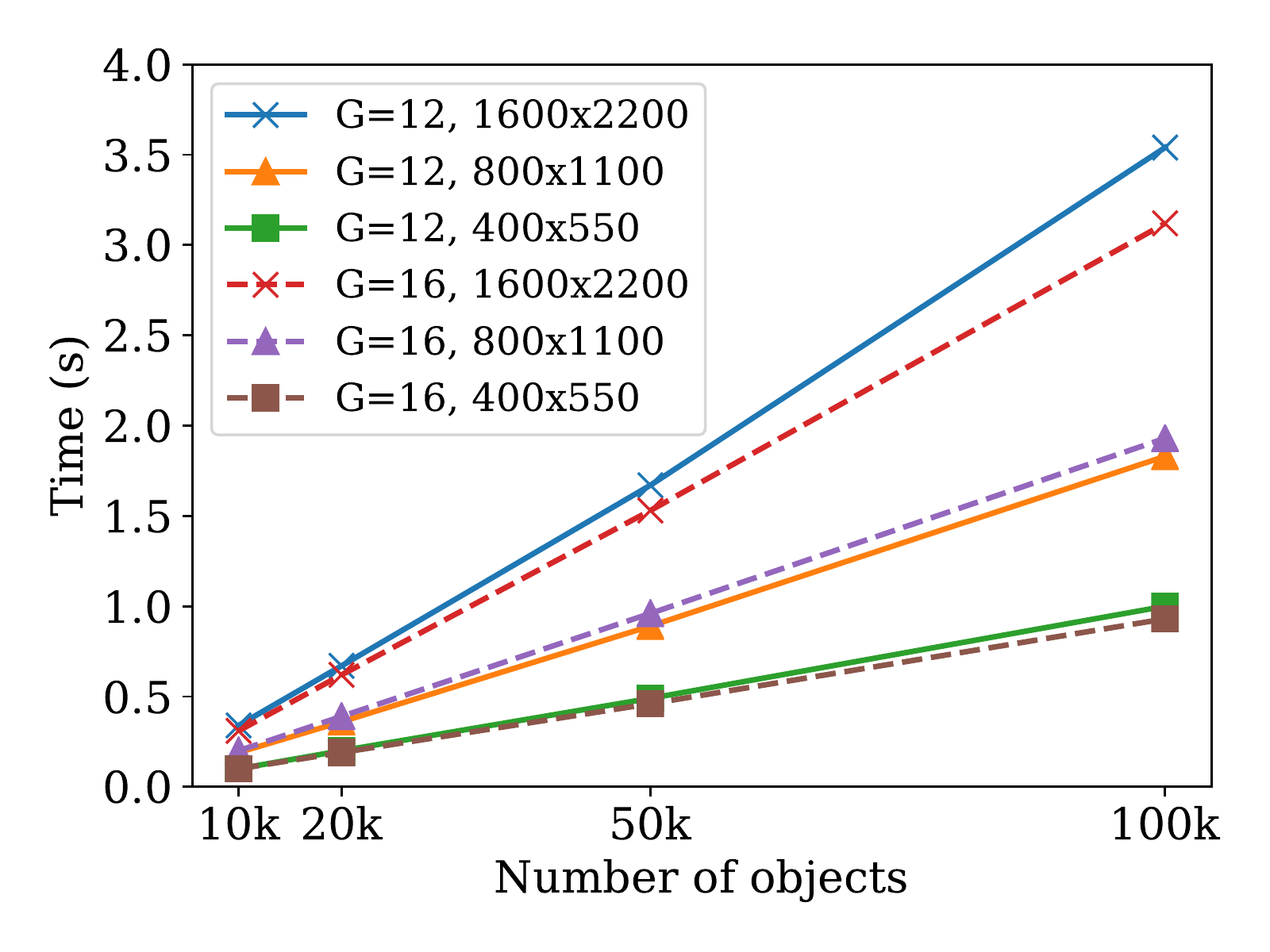}
		\label{fig:hxt_avg_query_time_nodes}  }
		\vspace{-5pt}
	\caption{HXT query performance}
	\label{fig:hxt_avg_query_time}
		\vspace{-15pt}
\end{figure}

\noindent
{\bf SSE Approach Evaluation.}
First, we measure the HXT index build time at the TC and the index size (Fig.~\ref{fig:hxt_index}). Each graph line corresponds to a combination of granularity ($L$) and maximum tree levels $h_{max}$. The index build time grows linearly with the number of objects. A finer granularity and a lower value of $h_{max}$ increase the build time, since they generate a larger number of keywords. 
For the $50k$ cardinality the index creation time never exceeds $20$ minutes, whereas in the worst case it takes $40$ minutes for the $100k$ case. For moderate granularity and height settings, build time is below $10$ minutes. The index size varies between $1$ and $5.5$GB.

\begin{figure}[tbh]
\vspace{-20pt}
	\subfloat[Keyword queries count]{
		\includegraphics[width=0.5 \columnwidth]{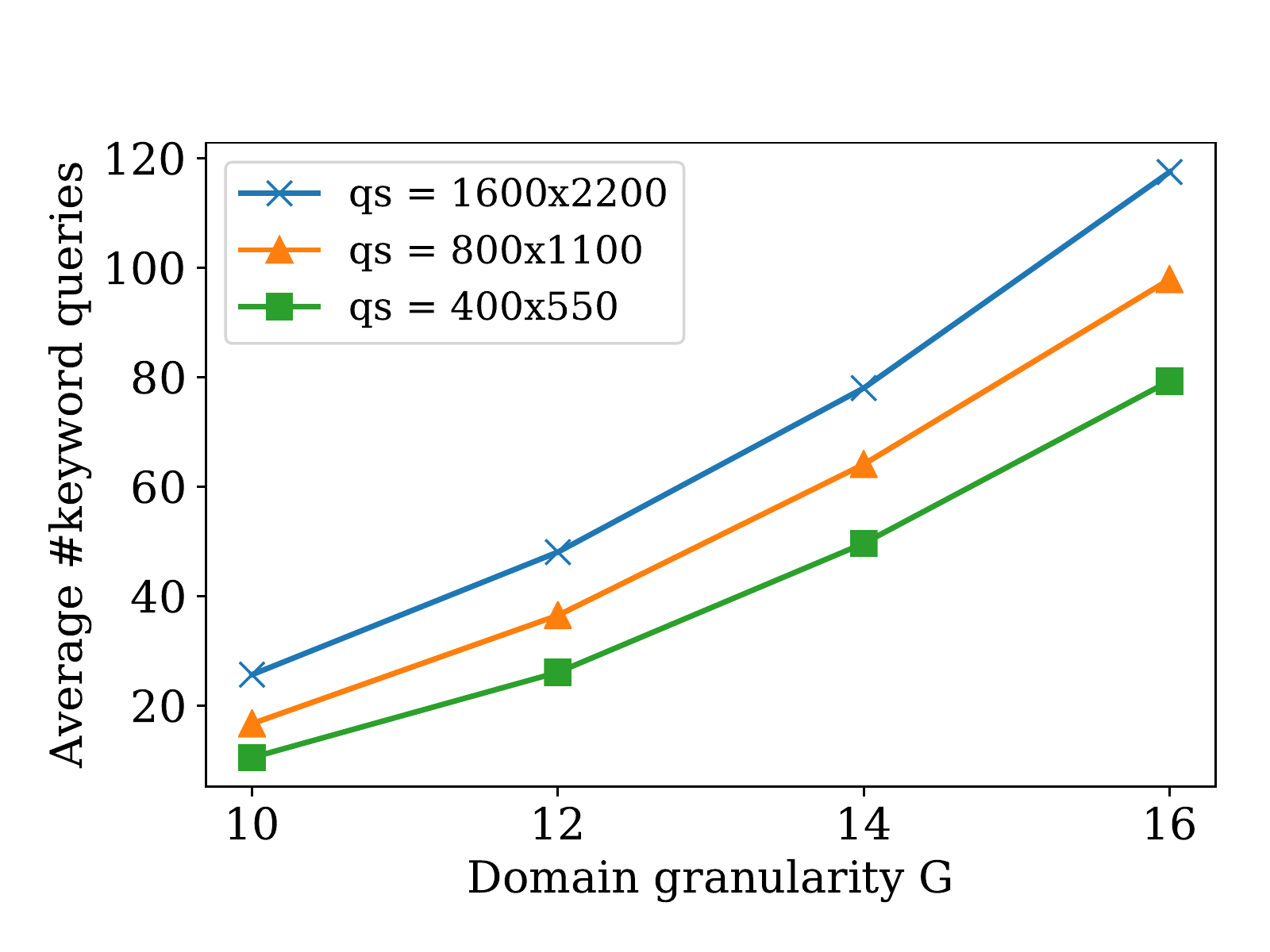}
		\label{fig:hxt_avg_keywords_queries_arbitrary}}
	\subfloat[\#Obj. containing first keyword]{
		\includegraphics[width=0.5 \columnwidth]{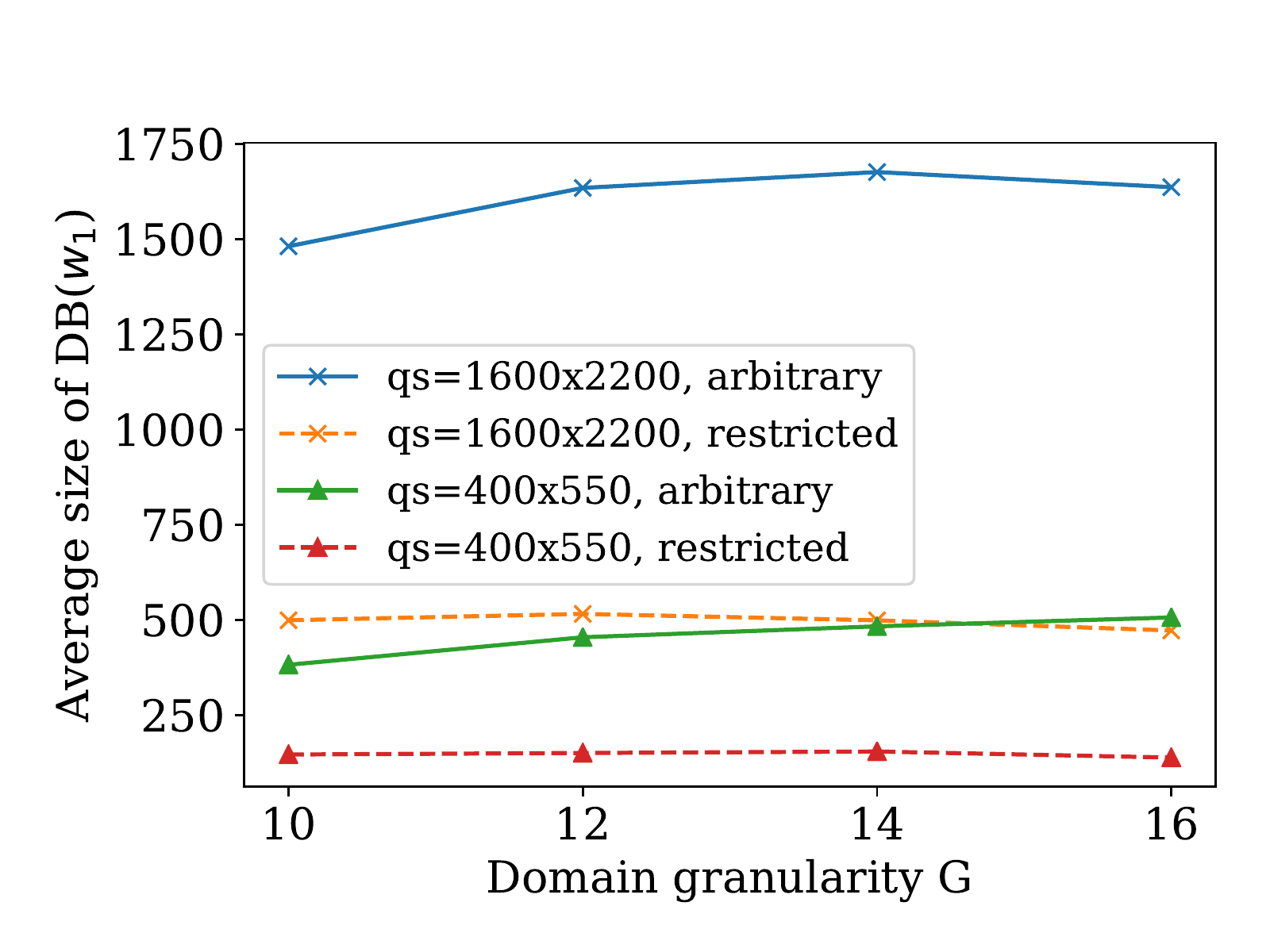}
		\label{fig:hxt_avg_t_set_stag_size_arbitrary}  }
			\vspace{-5pt}
	\caption{Analysis of query restriction effect}
	\label{fig:hxt_avg_keywords_queries_and_t_set_stag_size}
			\vspace{-10pt}
\end{figure}

Fig.~\ref{fig:hxt_avg_query_time_arbitrary} shows the average query time for arbitrarily placed queries.
The performance overhead is considerably higher for the queries with larger span and is less influenced by granularity. 
In the worst case, a query takes $12sec$, and if we exclude the largest query range, less than $6sec$ for $100k$ objects.
Fig.~\ref{fig:hxt_avg_query_time_nodes} shows the results when restricting query placement. 
Clearly, there is a significant gain in performance, resulting in a query time reduction between $2.5$ and $8$ times. 
The query time is always below $4sec$.
To better understand the performance gain due to query restrictions, we measure the count of individual conjunctive queries resulting from the decomposition of arbitrary ranges (Fig.~\ref{fig:hxt_avg_keywords_queries_arbitrary}). A single range may be decomposed into as many as $120$ conjunctive HXT queries, and the number of such queries increases with $G$. As a result, there is a much higher number of documents in the database containing the first keywords ($DB(w_1)$).
Fig.~\ref{fig:hxt_avg_t_set_stag_size_arbitrary} shows the cardinality of $DB(w_1)$, which increases significantly with the span of the query range. %

\begin{figure}
	\subfloat[Ciphertext generation time]{
		\includegraphics[width=0.5 \columnwidth]{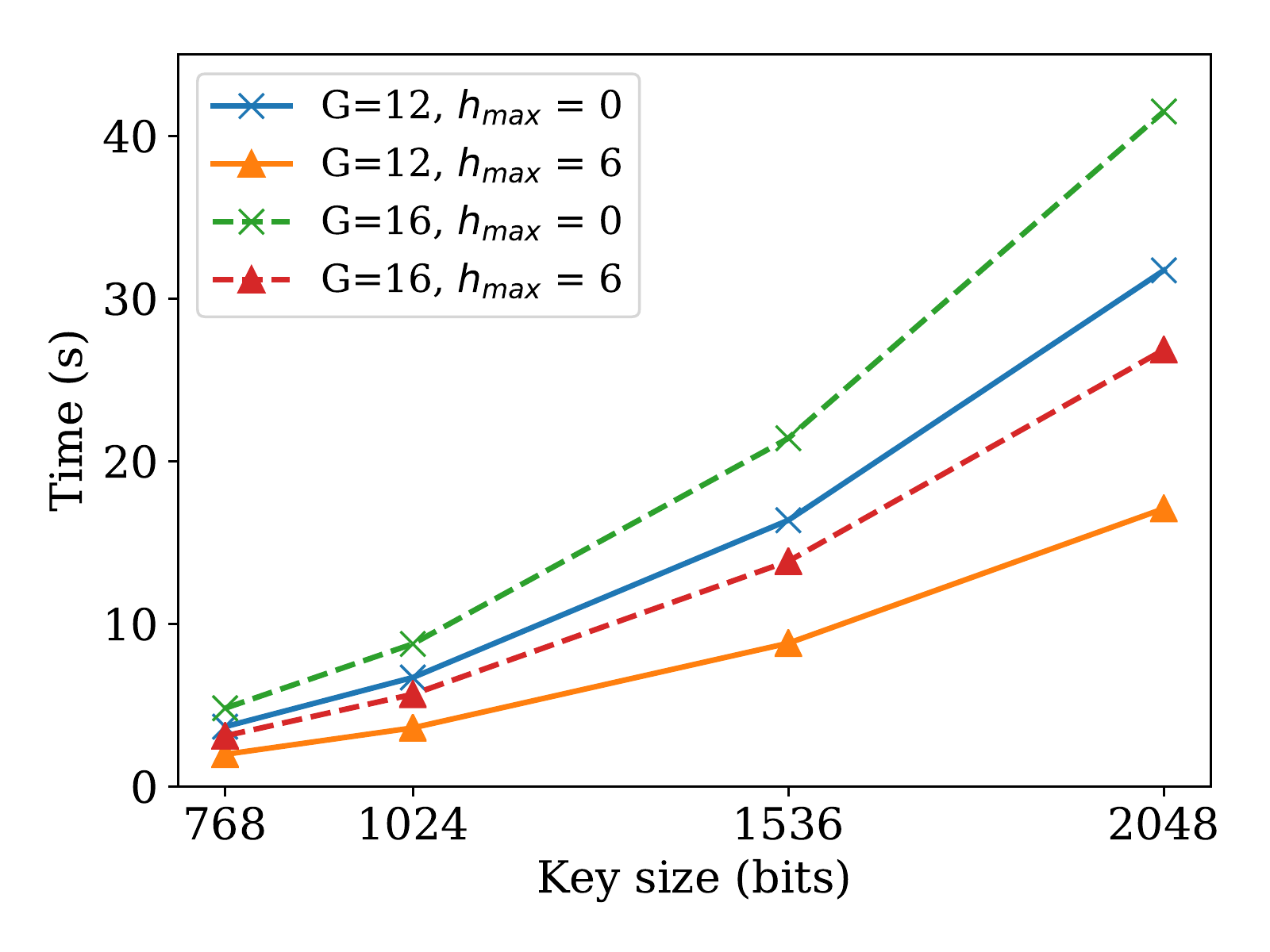}
		\label{fig:hve_cipher_time_per_item}  }
	\subfloat[Ciphertext size]{
		\includegraphics[width=0.5 \columnwidth]{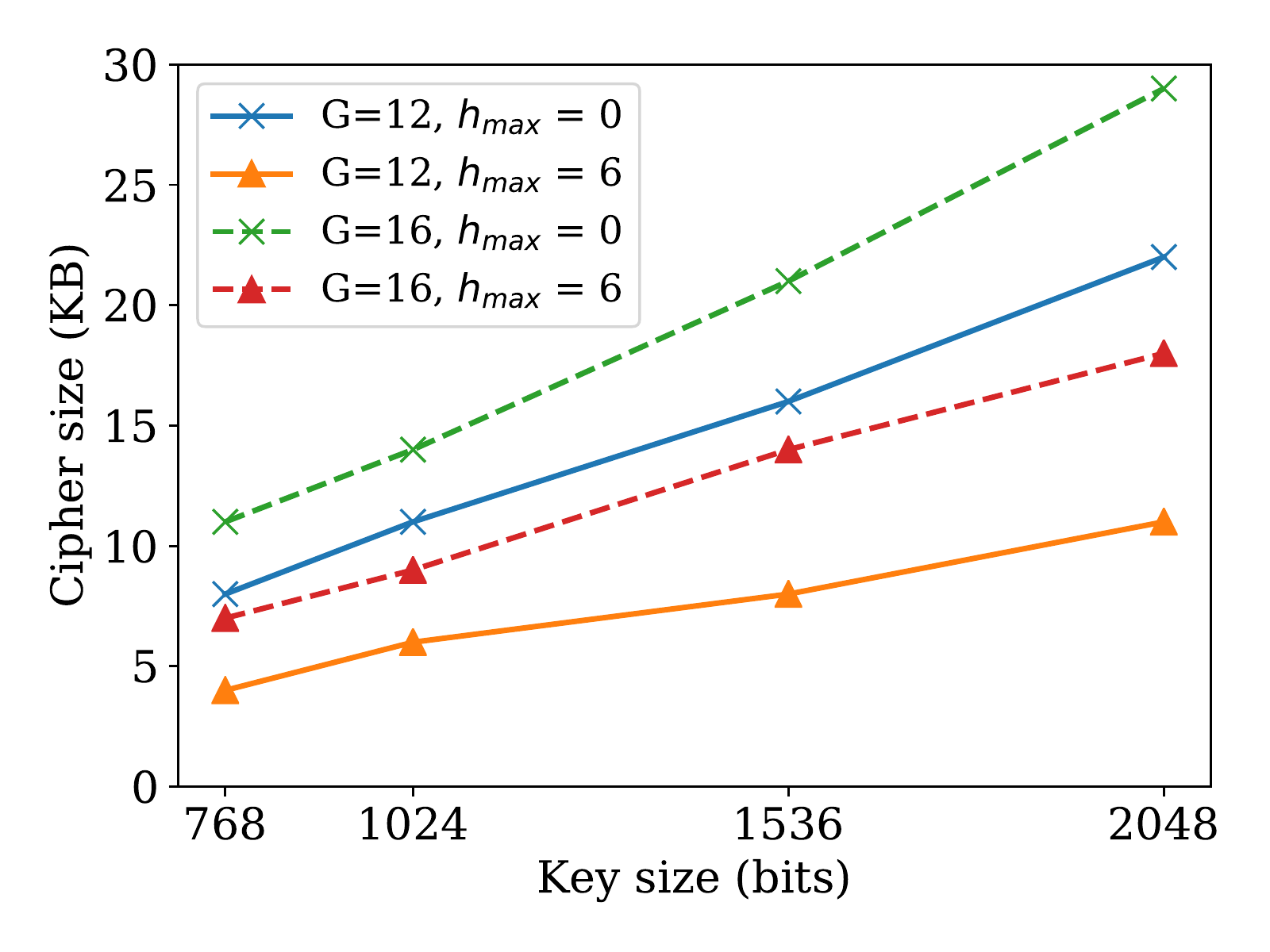}
		\label{fig:hve_cipher_size_per_item}}
			\vspace{-5pt}
	\caption{HVE encryption time and ciphertext size per object}
	\label{fig:hve_cipher_per_item}
	\vspace{-10pt}
\end{figure}

\noindent
{\bf HVE Approach Evaluation.}
With HVE, owners encrypt data by themselves, hence security is improved since locations are not shared with a trusted entity. However, the owner's device may have less computational power, so we must ensure the client overhead is low. Fig.~\ref{fig:hve_cipher_per_item} shows the HVE ciphertext generation time and size.
Current security guidelines specify $1024$-bit protection as sufficient for individual data; in this setting, encryption time is usually below $5sec$. Even for higher security requirements, encryption never exceeds $40sec$. Ciphertext size is under $30$KB.

\begin{table}[tbh]
	\centering
	\begin{tabular}{l|r|r|r|r}
		Key size            & 768   & 1024  & 1536  & 2048  \\ \hline \hline
		Generation time (s) & 0.019 & 0.036 & 0.085 & 0.165 \\ \hline
		Size (bytes)        & 402   & 534   & 786  & 1050
	\end{tabular}
	\caption{HVE token generation time and size}
	\vspace{-25pt}
	\label{tbl:hve_token}
\end{table}

Table~\ref{tbl:hve_token} shows the token generation time at the TA. 
Token generation time is short, $0.2sec$ in the worst case. Token size of is also negligible (at most $1$KB). The results justify our claim that there is a strong business case for TAs (e.g., cell operators) to participate in the system: the overhead is small, and no significant infrastructure investment is necessary to support such a service.

\begin{figure}
	\subfloat[Match time vs key length]{
		\includegraphics[width=0.5 \columnwidth]{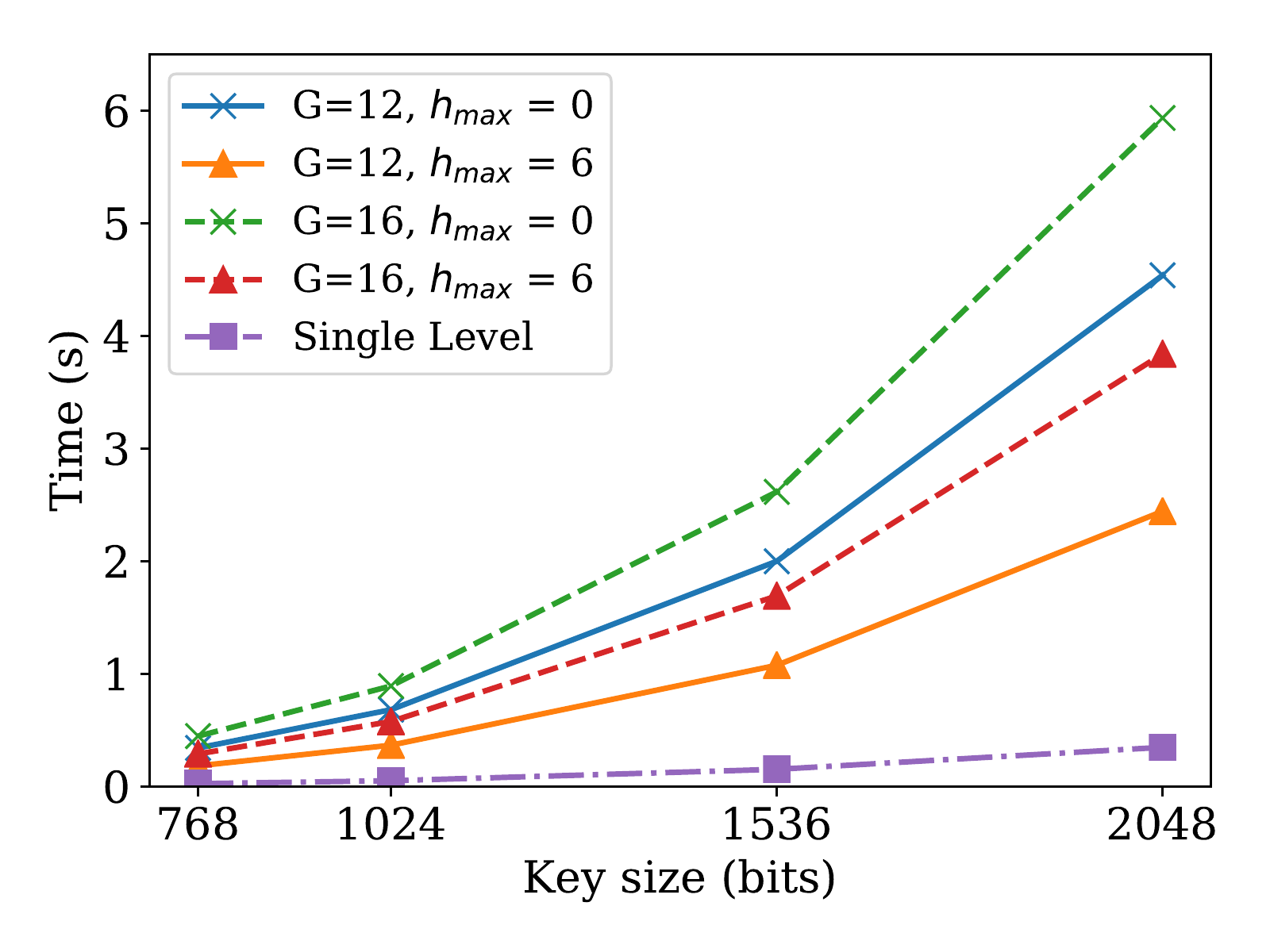}
		\label{fig:hve_query_time_vs_k}  }
	\subfloat[Match time vs domain granularity]{
		\includegraphics[width=0.5 \columnwidth]{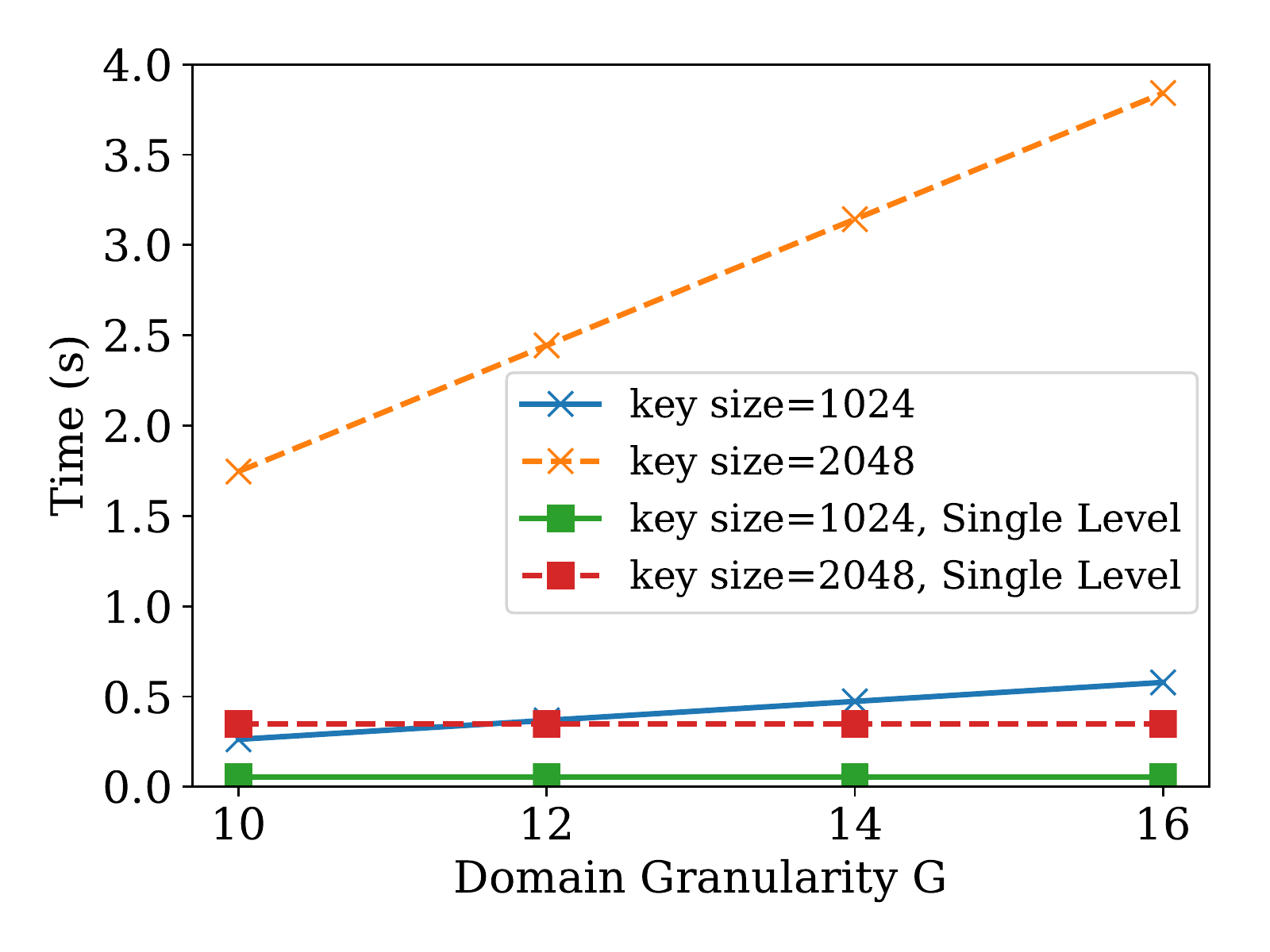}
		\label{fig:hve_query_time_vs_G}}
			\vspace{-10pt}
	\caption{HVE matching time per ciphertext}
	\label{fig:hve_avg_query_time_nodes}  
	\vspace{-10pt}
\end{figure}

The most significant performance concern with HVE is the query time, due to the use of expensive bilinear pairing operations. Fig.~\ref{fig:hve_avg_query_time_nodes} shows query time (i.e., HVE {\em match}) per token/ciphertext pair.
Query time for $1024$-bit security, $G=12$ (approximately $10$ meters per grid cell side) and query size is limited to $h_{\max} = 6$, is $370msec$, and reaching close to $6sec$ for higher security requirements (i.e. key size = 2048), larger $G = 16$, and no query size limit (i.e. $h_{\max} = 0$). 
With the {\em SingleLevel} optimization (Section~\ref{sec:hvetechnical}), the query time is reduced to under $52msec$. This is still significant though, considering that the marketplace may have millions of objects.

HVE is also expensive due to the absence of an index: every ciphertext is matched against the token. However, the search is embarrassingly parallel. One can distribute the ciphertexts over many nodes and achieve near-linear speedup. Since token size is small, it can easily be broadcast at query time to many nodes. Each Swarm node can perform the matching itself, and charge the buyer for the compute cycles.
Table~\ref{tbl:hve_parallel} shows parallel execution time when running HVE search for $10000$ objects on four cores, key size $1024$ bits, $G=16$, and $h_{\max} = 7$, with the {\em SingleLevel} optimization.
The speedup obtained is linear. Assuming linear speedup, 
using $100$ cores one could execute the query for a database of $10,000$ objects in $10$ seconds. Using a conservative cost estimation from cloud providers, this translates to a financial cost $\$0.03$/query, easily absorbed in the cost of a single marketplace transaction.

\begin{table}[tbh]
	\centering
	\begin{tabular}{l|r|r|r|r}
		\#processes            & 1   & 2  & 3  & 4  \\ \hline \hline
		Query time (s) & 525 & 332 & 221 & 170 \\ \hline
		Speedup        & 1   & 1.6   & 2.4  & 3.1
	\end{tabular}
	\caption{Parallel execution speedup for HVE matching}
	\vspace{-15pt}
	\label{tbl:hve_parallel}
\end{table}

\noindent
{\bf Financial Cost Evaluation.}
Our aim is to prove the financial viability of the proposed marketplace, by showing it represents only a small percentage of transaction cost. Financial cost includes the cost of writing to blockchain, and the cost of storage (index and metadata). 
We omit commitment computation time (which was under $0.2sec$).%
For blockchain operations, we measure both the gas and USD amount (at the rate of $1ether=\$133$).

\begin{table}[tbh]
	\centering
	\begin{tabular}{l|r|r}
		On-chain operation            & Gas   &  Cost (\$)  \\ \hline \hline
		Owner registration & 42150 & 0.02  \\ \hline
		Owner sets commitment public params        & 327590   & 0.10    \\ \hline
		TA/TC submits SSE/HVE index/file info        & 177160   & 0.06     \\ \hline
	\end{tabular}
	\caption{On-chain cost of one-time, setup operations}
	\vspace{-15pt}
	\label{tbl:eth_onetime_cost}
\end{table}

Table~\ref{tbl:eth_onetime_cost} shows the one-time set-up cost when joining the marketplace. This includes owner registration, owner setting up public parameters for digital commitments, and TA/TC submitting bootstrap information about SSE index or HVE flat file. We use $1024$ key size for digital commitments; the message space of each commitment is 64 bits (32 bits for each coordinate). 
The total set-up cost for each owner is about \$0.12, and for each TA in the HVE-based marketplace is \$0.06.

\begin{table}[tbh]
	\centering
	\begin{tabular}{l|r|r}
		On-chain operation            & Gas   &  Cost (\$)  \\ \hline \hline
		Owner submits commitment & 83092 & 0.02  \\ \hline
		Buyer makes an offer  & 297478   & 0.08 \\ \hline
		Owner withdraws payment  & 40649   & 0.01
	\end{tabular}
	\caption{On-chain cost breakdown for every transaction}
	\vspace{-25pt}
	\label{tbl:eth_cost}
\end{table}

Table~\ref{tbl:eth_cost} shows the cost of on-chain operations for an end-to-end purchase transaction between owner and buyer: an owner submits commitments for locations of her objects; 
a buyer makes an offer to purchase objects;
and optionally, the buyer withdraws his payment in case of discovering a fraudulent advertisement although this last step rarely occurs.
The total cost for a purchase is approximately \$0.11. This is a relatively low cost: considering an average of $\$10$ price per purchase, this represents a $1.1\%$ fee.

\vspace{-5pt}
\section{Related Work}
\label{sec:relatedwork}

Blockchain has been adopted in many areas such as healthcare~\cite{Ji2018BMPLS,kuo2018modelchain}, Internet of Things~\cite{FerragBlockchainIoT,kumar2018blockchain}, smart vehicles~\cite{Li2018CreditCoin,Knirsch2018PrivacyEV}, real-world asset trading~\cite{notheisen2017trading}, finance~\cite{hyvarinen2017blockchain}, or logistics~\cite{dobrovnik2018blockchain}. 
ModelChain~\cite{kuo2018modelchain} is a decentralized framework for privacy-preserving	healthcare predictive modeling based on a private blockchain. CreditCoin~\cite{Li2018CreditCoin} uses the blockchain and threshold ring signatures to achieve anonymity for smart vehicles. 
The work in~\cite{Ji2018BMPLS} proposed a blockchain-based location sharing scheme for telecare medical information systems. However, in their system, locations are encrypted using an order-preserving encryption scheme, which is known to incur significant leakage. Closer to our work,~\cite{Knirsch2018PrivacyEV} uses the blockchain-based model to protect locations of smart vehicles; however, their privacy model relies on random identifiers and enlargement of reported areas providing only ad-hoc protection, as opposed to our solution that inherits the strong protection of encryption.
In the context of data exchange and marketplaces for location data using blockchain, \cite{zyskind2015decentralizing} proposed a system based on conventional encryption, which does not support search on ciphertexts. Fysical~\cite{fysical} is a blockchain-based marketplace where suppliers can sell plaintext or aggregated location data, which raises serious privacy issues.

There are some important lines of works orthogonal to our approach.
One direction focuses on creating proofs of location using blockchain~\cite{locationproof-BrambillaAZ16,foam,platin,xyo}. 
The recently-proposed Hawk~\cite{kosba2016hawk} system is a blockchain model that provides transactional privacy such that private bids and financial data are hidden from public view. 
Zhang et. al~\cite{zhang2019gem2} proposed $GEM^2$-tree for authenticated range queries with off-chain storage.
Such approaches can be integrated into our system to allow validation of the geo-tags, hiding transactions, or validate results from our searchable encrypted indices.

\vspace{-5pt}
\section{Conclusion}
\label{sec:conclusion}

We proposed a blockchain-based privacy-preserving, accountable and spam-resilient marketplace for geospatial data which allows owners and buyers to be matched using only encrypted location information. To the best of our knowledge, this is the first approach to achieve these important desiderata. In future work, we will investigate architectures that eliminate the need for trusted entities or reduce the amount of information that is made available to such entities. In addition, we will explore alternative data encodings and encrypted processing techniques to further reduce system overhead. We will also investigate how our results can be extended to other types of data, not only geo-spatial attributes.

\textbf{Acknowledgments.} This research has been funded in part by NSF grants IIS-1910950 and  IIS-1909806, the USC Integrated Media Systems Center (IMSC), and unrestricted cash gifts from Google. Any opinions, findings, and conclusions or recommendations expressed in this material are those of the author(s) and do not necessarily reflect the views of any of the sponsors such as the NSF.

\bibliographystyle{ACM-Reference-Format}
\scriptsize
\bibliography{references}

\end{document}